\documentclass[11pt, a4paper]{article}
\pdfoutput=1

\usepackage[top=12mm,bottom=12mm,left=30mm,right=30mm,head=6mm,includeheadfoot]{geometry}
\usepackage{tocloft}

\usepackage[nottoc,notlot,notlof]{tocbibind}

\usepackage{titlesec}
\titlespacing*{\section}{0pt}{1.8\baselineskip}{\baselineskip}

\usepackage{fancyhdr}
\makeatletter
  \let\ps@plain\ps@fancy
\makeatother

\usepackage[utf8]{inputenc}
\usepackage{doi}
\usepackage{hyperref}
\usepackage{graphicx}

\usepackage{amsmath,amssymb,amsfonts,mathrsfs}
\usepackage{dsfont}
\usepackage[amsmath,thmmarks]{ntheorem}
\usepackage{tikz}

\usepackage{subcaption}

\usepackage{defs}

\usepackage{cite}

\begin{document}

\begin{center}{\Large \textbf{The Two-Particle Irreducible Effective Action for Classical Stochastic Processes}}\end{center}

\begin{center}
Tim Bode\textsuperscript{1*}
\end{center}

\begin{center}
{\bf 1} {German Aerospace Center (DLR), Linder H\"ohe, 51147 Cologne, Germany}
\\
* tim.bode@dlr.de
\end{center}

\noindent{}\textbf{Abstract}. By combining the two-particle-irreducible (2PI) effective action common in non-equilibrium quantum field theory with the classical Martin-Siggia-Rose formalism, self-consistent equations of motion for the first and second cumulants of non-linear classical stochastic processes are constructed. Such dynamical equations for correlation and response functions are important in describing non-equilibrium systems, where equilibrium fluctuation-dissipation relations are unavailable. The method allows to evolve stochastic systems from arbitrary Gaussian initial conditions. In the non-linear case, it is found that the resulting integro-differential equations can be solved with considerably reduced computational effort compared to state-of-the-art stochastic Runge-Kutta methods. The details of the method are illustrated by several physical examples.

\vspace{10pt}
\noindent\rule{\textwidth}{1pt}
\tableofcontents\thispagestyle{fancy}
\noindent\rule{\textwidth}{1pt}
\vspace{10pt}

\section{Introduction}

Time-dependent stochastic phenomena are ubiquitous in nature, ranging from the classic example of a single Brownian particle \cite{gardiner1985handbook, VanKampen2007} to the collective behavior of neurons \cite{stapmanns2020self} and financial markets \cite{black1973pricing, heston1993closed}. In complex systems such as neuronal networks, the mathematical description in terms of stochastic processes is complicated by the presence of non-linearities. In the absence of analytical methods to solve the corresponding non-linear stochastic differential equations (SDEs), one has to resort to numerical simulation or to approximate solutions. As has only recently been discussed in the context of neuronal networks \cite{stapmanns2020self, helias2019statistical}, the effective action \cite{peskin2018introduction}, which is well-known from quantum field theory (QFT), can in principle be used to derive perturbative expansions of non-linear stochastic processes to arbitrary order. The effective action is defined as the Legendre transform of the cumulant-generating functional with respect to the first cumulant and, as such, it is by construction restricted to self-consistency in the mean \cite{stapmanns2020self}. Being thus limited to calculating Feynman diagrams in terms of bare propagators, i.e.\ second cumulants without self-consistent corrections \cite{hertz2016path}, infinite resummations of classes of diagrams beyond the standard Dyson series are difficult to perform. For non-equilibrium dynamics, however, such advanced resummations turn out to be necessary to avoid problems of secularity and to obtain a solution with a uniform error scaling for all times \cite{Berges2004}. Moreover, treating also the correlation and response functions as ``order parameters'' {is} particularly useful out of equilibrium, where the correlations can no longer be determined via the linear response of the mean \cite{hertz2016path}. {The general problem of finding self-consistent solutions for the first \textit{and} second cumulant is, of course, equally important both in and out of equilibrium, as shown by the work of De Dominicis \cite{DeDominicis1978}, who uses the second Legendre transform on disordered systems including Gaussian white noise to derive equilibrium properties.}

In non-equilibrium QFT, the two-particle irreducible (2PI) effective action \cite{Cornwall1974, Berges2004, Berges2015, calzetta2008nonequilibrium}, where an additional Legendre transform with respect to the second moment is employed, has proven to be a powerful tool to derive self-consistent equations with the above properties. It has found, however, little discussion in the context of classical stochastic processes. In this work, we give a detailed discussion of the 2PI effective action as it applies to noise-driven classical systems {out of equilibrium}.

Perturbative expansions involving Feynman diagrams with resummed propagator lines have been discussed in relation to stochastic processes in Ref.\ \cite{hertz2016path}, based on heuristically replacing bare by resummed propagator lines in the self-energy diagrams. The 2PI framework now represents a rigorous and well-defined approach to self-consistent resummation, which in combination with the Martin-Siggia-Rose (MSR) formalism is an important step to improve the field-theoretic description of non-linear stochastic processes. Note that a related method has been proposed as an extension of the \textit{Plefka expansion} from discrete to continuous degrees of freedom in Ref.\ \cite{Bravi_2016}, where, however, it is applied to linear stochastic processes only. We show here that, in particular, the treatment of \textit{non-linear} stochastic processes follows more naturally by combining the concepts of the 2PI effective action with the MSR formalism, and opens up the entire toolbox of QFT for treating such non-linearities. { The 2PI construction has also been applied to Doi-Peliti field theory \cite{Bothe2021} in order to study the colloidal glass transition in equilibrium \cite{Jacquin2011}. The present work can be seen as a generalization of the methods outlined in Ref.\ \cite{Jacquin2011} to \textit{non-equilibrium}, in particular as the diagrammatic expansion is presented here in general terms and thus not limited to the MSR formalism.}

While all effective actions are equivalent if no approximations are made, they do in general differ once the perturbation series is truncated at a given order. Thus at the same formal order in the coupling, the 2PI effective action resums a greater set of diagrams than the singly Legendre-transformed effective action, which leads to noticeable corrections already for moderate non-linearities (cf. section \ref{subsec:quad_non_lin}). The 2PI effective action furthermore allows for the {convenient} investigation of the system dynamics starting from \textit{arbitrary} Gaussian initial states \cite{Berges2015}. {Note that general $n$-particle irreducible ($n$PI) effective actions up to $n=4$ are discussed at length in Refs.\ \cite{doi:10.1063/1.1704062, vasiliev1998functional}}.

This paper is organized as follows. In section \ref{subsec:first_legendre} we perform the first Legendre transform on the cumulant-generating functional and obtain the resulting effective action. This is followed in section \ref{subsec:second_Legendre} by the second Legendre transform, which leads us to the 2PI effective action. As a simple illustration, we discuss zero-dimensional $\phi^4$ theory. In section \ref{subsec:MSR}, we introduce the MSR formalism as our framework of choice for treating SDEs. These techniques are then applied to the Ornstein-Uhlenbeck process in section \ref{subsec:OU_process}. In section \ref{subsec:geom_brownian}, we show how to obtain analytical results for geometric Brownian motion with this method. Finally, in section \ref{subsec:quad_non_lin} we give a detailed analysis of SDEs with additive noise and quadratic non-linearity. Section \ref{sec:conclusion} then concludes the paper with final remarks.

\section{Effective-Action Formalism}

A stochastic process $x(t)$ can be described by the SDE
\myEq{\label{eq:SDE}
    \dd x(t) = f[x(t)]\dd t + \sigma[x(t)]\dd W(t),
}
where $W(t > 0)$ is a Brownian motion in one dimension. Alternatively, the stochastic process can be represented via the Onsager-Machlup path integral \cite{onsager1953fluctuations}
\myEq{\label{eq:OM_path_integral}
    \int\mathcal{D}x\exp{\cbs{-\frac{1}{2}\mint{0}{t_f}{t}\rbs{\frac{\partial_t{x}(t) - f[x(t)]}{\sigma[x(t)]}}^2}},
}
which follows after discretizing the path $x(t)$ into $N$ segments such that $x(0) = x_0$ and $x(t_f) = x_N$, this endpoint being excluded from the functional integration. Then Eq.\ \eqref{eq:OM_path_integral} is proportional to the probability to find the system at $x_N$ after time $t_f$. By performing a Hubbard-Stratonovich transformation, it is possible to recast expression \eqref{eq:OM_path_integral} as $\int\mathcal{D}x\mathcal{D}z\, \ee^{-S[x, z]}$, where 
\myEq{\label{eq:OM_Hubbard_Straonovich}
    S[x, z] = \mint{0}{t_f}{t}\cbs{\sigma^2[x(t)] z^2(t)/2 + \ii z(t)\rbs{\partial_t{x}(t) - f[x(t)]}}
}
is the corresponding action functional. Observe that we employ It\^o discretization here and throughout \cite{gardiner1985handbook}. For detailed discussions of the functional measures and other discretization conventions, consider Refs.\ \cite{Cugliandolo_2019, Cugliandolo_2017, hertz2016path, Hanggi1989}. Historically, the formalism behind Eq.\ \eqref{eq:OM_Hubbard_Straonovich} was pioneered by Martin, Siggia and Rose in operator form \cite{Martin1973}. Later, its path-integral representation was given by Janssen \cite{janssen1976lagrangean} and De Dominicis \cite{DeDomnicis1978}. {Some applications of the MSR formalism to disordered systems can be found in Refs.\ \cite{PhysRevE.68.046101, PhysRevE.69.061107}.} Compared to the Onsager-Machlup path integral, it has certain formal advantages that will be touched upon below. It is also equivalent to the Keldysh path integral of bosonic open quantum systems that admit proper probability distributions on phase space \cite{Sieberer2015}.

\subsection{First Legendre Transform}\label{subsec:first_legendre}

Let us generalize Eq.\ \eqref{eq:OM_Hubbard_Straonovich} to an action functional comprising an arbitrary number of time-dependent fields $\phi_a(t)$ with indices $a=1, ..., n$, denoted collectively by $\bs{\phi}(t)$. Then we can define a moment-generating or partition functional as
\myEq{\label{eq:MGF}
    Z[\bs{j}] = \left\langle \exp\rbs{\mint{0}{t_f}{t}\bs{j}^T(t)\bs{\phi}(t)/c} \right\rangle =  \int\mathcal{D}\bs{\phi}\;\exp\rbs{\mint{0}{t_f}{t}\bs{j}^T(t)\bs{\phi}(t)/c - S[\bs{\phi}]/c},
}
where we have introduced a constant $c$ that formally plays the same role as $\hbar$ in QFT, and $\bs{j}(t)$ is the source field. The moment-generating functional $Z$, in turn, gives rise to the \textit{scaled} cumulant-generating function (CGF)
\myEq{
    \mathcal{W}[\bs{j}] = c\ln Z[\bs{j}],
}
where formally one should take $c\to 0$ \cite{touchette2009large}. However, since we are usually interested in the case of finite $c^{-1}$, we do not include this limit into the definition of $\mathcal{W}(j)$. In QFT, the analogous procedure is to let $\hbar\to 0$, which leads to the standard loop expansion in the limit of small quantum fluctuations \cite{peskin2018introduction}.

For asymptotically small $c$, the path integral in Eq.\ \eqref{eq:MGF} becomes very strongly peaked, such that its main contribution comes from the saddle point, and one may approximate
\myEq{\label{eq:SP_approx}
        \mathcal{W}[\bs{j}] \approx \sup_{\bs{{\phi}}}\rbs{\mint{0}{t_f}{t}\bs{j}^T(t)\bs{{\phi}}(t) - S[\bs{{\phi}}]}.
}
Now Eq.\ \eqref{eq:SP_approx} suggests \textit{defining} the so-called one-particle irreducible (1PI) effective action $\Gamma$ as the Legendre transform \cite{touchette2009large} of the scaled CGF with respect to the first cumulant:
\myEq{\label{eq:def_Gamma}
    \Gamma[\bs{\bar{\phi}}] := \sup_{\bs{j}}\rbs{\mint{0}{t_f}{t}\bs{j}^T(t)\bs{\bar{\phi}}(t) - \mathcal{W}[\bs{j}]}.
}
{When} the scaled CGF is expanded into tree graphs where the functional derivatives of $\Gamma[\bs{\bar{\phi}}]$ act as vertices connected by edges $\delta^2 \mathcal{W}[\bs{j}]/\delta j_a(t)\delta j_b(t')$, then $\Gamma[\bs{\bar{\phi}}]$ is found to contain no graphs that can be disconnected by cutting a single edge \cite{helias2019statistical}{, for which reason it is called ``one-particle irreducible''. }

The 1PI effective action $\Gamma[\bs{\bar{\phi}}]$ is generally not identical to $S[\bs{\bar{\phi}}]$, as one can see also from the fact that the classical saddle-point $\bs{\phi}_c^*$  defined by $ 0 = \myFrac{\delta S[\bs{\phi}_c^*]}{\delta\bs{\phi}_c^*(t)}$ (i.e. neglecting fluctuations) may deviate from the true expectation value $\E{\bs{\phi}(t)}$, e.g.\ when $S$ is not Gaussian and $c$ is finite. We shall assume that the supremum in Eq.\ \eqref{eq:def_Gamma} always exists (and is unique), such that we can calculate the Legendre transform as usual by finding the source $\bs{j}(t)$ as a function of $\bs{\bar{\phi}}(t)$ from
\myEq{
        \bs{\bar{\phi}}(t) = \myFrac{\delta \mathcal{W}[\bs{j}]}{\delta\bs{j}(t)},
}
the inverse relation to which is
\myEq{\label{eq:1PI_inv_Legendre}
    \bs{j}(t) = \myFrac{\delta \Gamma[\bs{\bar{\phi}}]}{\delta\bs{\bar{\phi}}(t)}.
}
After setting the source field on the left-hand side of Eq.\ \eqref{eq:1PI_inv_Legendre} to zero, its solution provides an approximation to $\E{\bs{\phi}(t)}$ that becomes exact as $c$ vanishes (in which case $\bs{\bar{\phi}}(t)$ becomes identical to $\bs{\phi}_c^*$) {or if $\Gamma[\bs{\bar{\phi}}]$ happens to be known exactly}. To first order in $c$ (one-loop order in QFT terms \cite{Berges2004, Berges2015}), the 1PI effective action reads
\myEq{\label{eq:1PIEA}
    \Gamma[\bs{\bar{\phi}}] = \mathrm{const.} + S[\bs{\bar{\phi}}] + \frac{c}{2}\operatorname{Tr}\ln \bs{{G}}_0^{-1}[\bs{\bar{\phi}}] + \mathcal{O}\rbs{c^2}, 
}
where $\bs{{G}}_0^{-1}[\bs{\bar{\phi}}]$ is the Hessian of the action $S$. For Gaussian action functionals $S$, the inverse Green function $\bs{{G}}_0^{-1}$ is independent of $\bs{\bar{\phi}}$ and Eq.\ \eqref{eq:1PIEA} is exact.

An instructive example is provided by zero-dimensional $\phi^4$ theory, for which an investigation of the \textit{quantum} (2PI) effective action can be found in Ref.\ \cite{millington2019visualising}. In the present classical case, we consider the action
\myEq{\label{eq:0D_action_phi_4}
    S(\phi) = \frac{\rbs{\phi - \mu}^2}{2\sigma^2} + \frac{g}{4!}\phi^4, 
}
which describes a normal distribution supplemented by a quartic non-linearity. Eq.\ \eqref{eq:1PIEA} then evaluates to
\myEq{
    \Gamma(\bar{\phi}) &=  \mathrm{const.} + S(\bar{\phi}) + \frac{c}{2}\ln\rbs{\sigma^{-2} + g\bar{\phi}^2/2} + \mathcal{O}\rbs{c^2},
}
where $G_0^{-1}(\phi) = \sigma^{-2} + g\bar{\phi}^2/2$. Setting $j(t)=0$ in the scalar version of Eq.\ \eqref{eq:1PI_inv_Legendre}, we obtain 
\myEq{\label{eq:0D_1PI_mf}
 0 &= \sigma^{-2}\rbs{\bar{\phi} - \mu} + \frac{g \bar{\phi}^3}{6} + \frac{cg\bar{\phi}}{2\rbs{\sigma^{-2} + \myFrac{g \bar{\phi}^2}{2}}}.
}
The fact that the probability density $\exp\rbs{-S(\phi) / c}$ concentrates around the ``mean-field'' value, which is the solution to Eq.\ \eqref{eq:0D_1PI_mf} for $c\to 0$, is the basic principle behind the tree-level approximation well-known from the field theory of statistical physics \cite{helias2019statistical}. The last term on the right-hand side of Eq.\ \eqref{eq:0D_1PI_mf} is the first correction beyond tree level and arises from quadratic fluctuations around the saddle point. Seen as a coupling expansion in $g$, moreover, solving this equation with a non-linear solver produces a uniform approximation \cite{Berges2004} to the exact result that is superior to a na\"\i ve expansion of the partition function.

\subsection{Second Legendre Transform}\label{subsec:second_Legendre}

\begin{figure}
	\centering
	\includegraphics[width=0.7\textwidth]{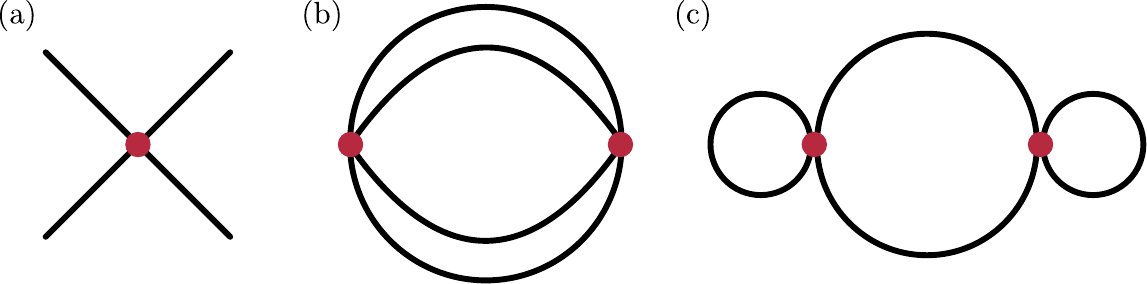}
	\caption{ Consider an action functional of the form $\left.S[\phi(t)] = \int \dd t\, S(\phi)\right|_{\phi = \phi(t)}$, where $S(\phi)$ was defined in Eq.\ \eqref{eq:0D_action_phi_4}. Then (a) is a graphical representation of the corresponding quartic vertex, which is proportional to the coupling $g$. In (b), a second-order two-particle irreducible (2PI) graph is shown.  The graph in (c) is of the same order in $g$, yet two-particle \textit{reducible} since it can be cut in two by cutting two lines.}
\label{fig:2PI_2PR}
\end{figure}

While the 1PI effective action in Eq.\ \eqref{eq:1PIEA} can in principle be used to extract information about cumulants of all orders by means of vertex functions \cite{stapmanns2020self}, self-consistency will by construction be given only for the first cumulant, i.e. the mean. It is, however, often desirable to also compute the \textit{second} cumulant (i.e. the variance) self-consistently. This is possible by extending the above construction via an additional Legendre transform with respect to the second \textit{moment} \cite{Cornwall1974, Berges2015}. For this purpose, we introduce a (symmetric) two-point source $\bs{K}(t, t')$ into the moment-generating functional. In terms of $\mathcal{W}$, we have
\begin{align*}
    \ee^{c^{-1} \mathcal{W}[\bs{j}, \bs{K}]} = \int\mathcal{D}\bs{\phi}\exp{\cbs{\frac{1}{c}\int_{0}^{t_f}{\hspace{-0.125cm}\dd t}\; \sbs{\bs{j}^T(t)\bs{\phi}(t) + \frac{1}{2} \int_{0}^{t_f}{\hspace{-0.125cm}\dd t'}\; \bs{\phi}^T(t)\bs{K}(t, t')\bs{\phi}(t')} - S[\bs{\phi}] / c}}.
\end{align*}
The doubly Legendre-transformed effective action is then defined as
\begin{align*}
    \Gamma[\bs{\bar{\phi}}, \bs{ G}] =  \int_{0}^{t_f}{\hspace{-0.15cm}\dd t} \sbs{ \bs{j}^T(t)\bs{\bar{\phi}}(t) + \frac{1}{2} \int_{0}^{t_f}{\hspace{-0.15cm}\dd t'}\, \operatorname{Tr}\bs{K}(t, t')\rbs{c\bs{ G}(t, t') + \bs{\bar{\phi}}(t)\bs{\bar{\phi}}^T(t')} } - \mathcal{W}[\bs{j}, \bs{K}],  
\end{align*}
where $\bs{\bar{\phi}}(t)$ and $c\bs{ G}(t, t')$ are the first and second cumulant in the presence of the external source fields $\bs{j}(t)$ and $\bs{K}(t, t')$, i.e.\
\myEq{\label{eq:2PI_legendre_vars}
    \bar{\phi}_{a}(t) &= \frac{\delta \mathcal{W}[\bs{j}, \bs{K}]}{\delta j_a(t) } , \\
     G_{ab}(t, t') &= \frac{\delta^2 \mathcal{W}[\bs{j}, \bs{K}]}{\delta j_a(t) \delta j_b(t')}.
}
Setting the source fields to zero, one recovers the expectation values
\myEq{
    \left.\frac{\delta \mathcal{W}[\bs{j}, \bs{K}]}{\delta j_a(t) } \right|_{\bs{j}, \bs{K} = 0} &= \E{\phi_a(t)}, \\
    \left.\frac{\delta^2 \mathcal{W}[\bs{j}, \bs{K}]}{\delta j_a(t) \delta j_b(t')} \right|_{\bs{j}, \bs{K} = 0} &= c^{-1}\sbs{\E{\phi_a(t)\phi_b(t')} - \E{\phi_a(t)}\E{\phi_b(t')}} .
}
The second moment can also be obtained via
\myEq{\label{eq:intro_second_moment}
    \left.\frac{\delta \mathcal{W}[\bs{j}, \bs{K}]}{\delta K_{ab}(t, t')}\right|_{\bs{j}, \bs{K} = 0} &= {\frac{1}{2}} \E{\phi_a(t)\phi_b(t')}.
}
{Note that it is also possible to interpret the functional derivative with respect to $K$ symmetrically, i.e.\ $\delta/\delta K_{ab} + \delta/\delta K_{ba}$, which removes the factor of $1/2$ in Eq.\ \eqref{eq:intro_second_moment} and the ensuing equations below.} In close connection to the 1PI effective action in Eq.\ \eqref{eq:1PIEA}, the 2PI effective action \cite{Berges2004, Berges2015} can now be shown to read
\myEq{\label{eq:2PIEA}
    \Gamma[\bs{\bar{\phi}}, \bs{ G}] = \mathrm{const.} + S[\bs{\bar{\phi}}] + \frac{c}{2}\operatorname{Tr}\ln \bs{ G}^{-1} + \frac{c}{2}\operatorname{Tr}\bs{G}_0^{-1}[\bs{\bar{\phi}}]\bs{ G} + \Gamma_2[\bs{\bar{\phi}}, \bs{ G}],
}
where $\Gamma_2[\bs{\bar{\phi}}, \bs{ G}] = \mathcal{O}(c^2)$ contains the non-linear contributions taken into account beyond the quadratic fluctuations around the saddle point. When $\Gamma_2[\bs{\bar{\phi}}, \bs{G}]$ is represented in terms of Feynman diagrams \cite{hertz2016path}, by construction there appear no diagrammatic graphs that can be cut in two by cutting two edges \cite{Berges2004}. Since the graph edges $G_{ab}(t, t')$ represent particle propagators in QFT, this property is called ``two-particle irreducibility``. Graphical examples of a 2PI diagram and a two-particle \textit{reducible} diagram are given in Fig.\ \ref{fig:2PI_2PR}(b) and (c), respectively. 

Note how $\Gamma$ {also depends} on $\bs{G}$: this is the hallmark of self-consistency for the second cumulant. While to first order in $c$, Eq.\ \eqref{eq:2PIEA} is equivalent to Eq.\ \eqref{eq:1PIEA}, any non-trivial $\Gamma_2$ now results in self-consistent corrections to both $\bs{\bar{\phi}}$ and $\bs{G}$. 
After setting the external sources to zero, these self-consistent solutions are determined via the {pair} of equations
\myEq{\label{eq:2PI_EoM}
    0 = \frac{\delta\Gamma[\bs{\bar{\phi}}, \bs{ G}]}{\delta \bar{\phi}_a(t)}, \qquad
    0 = \frac{\delta\Gamma[\bs{\bar{\phi}}, \bs{ G}]}{\delta  G_{ba}(t', t)},
}
the first of which gives the equation of motion of the fields $\bar{\phi}_a(t)$, while the second may be written as
\myEq{\label{eq:dyson}
    \frac{c}{2} G^{-1}_{ab}(t, t') = \frac{c}{2} G_{0,\, ab}^{-1}(t, t') + \frac{\delta \Gamma_2}{\delta  G_{ba}(t', t)},
}
which we recognize as Dyson's equation upon defining the so-called \textit{self-energy}
\myEq{\label{eq:2PI_self_energy}
    \Sigma_{ab}(t, t') = -2 c^{-1} \frac{\delta \Gamma_2}{\delta  G_{ba}(t', t)},
}
which can be a functional of both $\bs{\bar{\phi}}$ and $\bs{ G}$. Now it is well-known that in terms of Feynman diagrams, the self-energy {does} not contain any graphs that can be cut in two by cutting a single propagator line $G_{ab}(t, t')$ (one-particle irreducibility). As $\Sigma_{ab}(t, t')$ is defined as the functional derivative of $\Gamma_2[\bs{\bar{\phi}}, \bs{ G}]$ in the present formalism, one can easily prove by contradiction \cite{Berges2004} that $\Gamma_2[\bs{\bar{\phi}}, \bs{ G}]$ must indeed be 2PI.

With $G_{0,\, ab}^{-1}(t, t') = \delta(t - t')G_{0,\, ab}^{-1}(t, t)$, Eq.\ \eqref{eq:dyson} is then also equivalent to an integro-differential equation of \textit{Kadanoff-Baym} type \cite{Baym1961}, i.e. 
\myEq{\label{eq:Kadanoff-Baym}
    \delta(t - t')\mathds{1} = \bs{G}_{0}^{-1}(t, t)\bs{G}(t, t') - \mint{0}{t_f}{s}{\bs{\Sigma}(t, s)}\bs{G}(s, t').
}
For illustration, we go back to the zero-dimensional $\phi^4$ action of Eq.\ \eqref{eq:0D_action_phi_4}. We can again perform a self-consistent expansion in the small parameter $c$ (the equivalent of the well-known loop expansion in $\hbar$). The new variables of the Legendre transform are now defined as
\myEq{
    \bar{\phi} = \frac{\delta \mathcal{W}}{\delta j}, \qquad
    \bar{\phi}^2 + c G = \frac{\delta \mathcal{W}}{\delta K/2}.
}
By developing $\bar{\phi}$ and $G$ in a formal power series, e.g.\ $\bar{\phi} = \bar{\phi}_0 + c \bar{\phi}_1 + c^2 \bar{\phi}_2 + \mathcal{O}(c^3)$, one can then show that up to second order in both $c$ and $g$  \cite{carrington20044pi}, the 2PI effective action is
\myEq{
    \Gamma(\bar{\phi}, G) &= \mathrm{const.} + S(\bar{\phi}) + \frac{c}{2}\left(\sigma^{-2} G + \ln G^{-1} + \myFrac{g \bar{\phi}^{2} G}{2} \right) \\
    &+ c^2 \left(\frac{g}{8}G^{2} - \frac{g^{2}}{12}\bar{\phi}^{2}G^{3}\right) + \mathcal{O}\rbs{c^3}.
}
Clearly, the {solutions for the two variables will no longer be independent.} This becomes particularly manifest from Eqs.\ \eqref{eq:2PI_EoM}, which in this case read
\begin{subequations}\label{eq:0D_2PI}
    \begin{align}
         0 &= \sigma^{-2}\rbs{\bar{\phi} - \mu} + \myFrac{g \bar{\phi}^3}{6} + cg\rbs{\bar{\phi} G - \myFrac{cg \bar{\phi} G^3}{3}} / 2, \label{eq:0D_2PI_mf} \\
         0 &= G^{-1} - \sigma^{-2} - \myFrac{g \bar{\phi}^2}{2} -  cg\rbs{G - g\bar{\phi}^2 G^2} / 2.  \label{eq:0D_2PI_G}      
    \end{align}
\end{subequations}
Comparing this to the one-loop result in Eq.\ \eqref{eq:0D_1PI_mf}, the main difference is the self-consistency in the second cumulant, which leads to a coupled set of non-linear equations. To first order in $c$, Eqs.\ \eqref{eq:0D_2PI} naturally reduce to Eq.\ \eqref{eq:0D_1PI_mf}.

\subsection{Martin-Siggia-Rose Formalism}\label{subsec:MSR}

A standard construction to find the path integral of a stochastic process consists of the following steps: The sought-for transition probability $p(x|x_0)$ from an earlier point $x_0$ to a later point $x$ is expressed as the marginal of the joint probability to find the system at a (fixed) number of intermediate points. When the distribution of the noise is assumed to be given, it is then possible to express $p(x|x_0)$ in terms of the characteristic functional of the noise \cite{Hanggi1989}, which requires knowledge of the corresponding Jacobian. In this work, It\^o regularization is assumed throughout for clarity, such that the Jacobian determinant becomes trivial (for details on this topic, see {\cite{gardiner1985handbook, hertz2016path}}). The construction of the path integral via the noise characteristic functional introduces a special structure of the corresponding field theory that can be expressed by letting
\myEq{
\label{eq:MSR_fields}
    \bs{\phi}(t) = 
        \begin{pmatrix}
            x(t) \\ z(t)
        \end{pmatrix}, \qquad 
        \bs{j}(t) = 
        \begin{pmatrix}
            h(t) \\ j(t)
        \end{pmatrix}, 
}
where $z(t)$ is the so-called \textit{response} field (often written $\hat{x}$), and we have also introduced the respective source fields $h(t)$ and $j(t)$. Note that the source field $\bs{K}(t, t')$ is now a $2es 2$ matrix acting on the MSR fields in Eq.\ \eqref{eq:MSR_fields}. The first cumulants are
\myEq{\label{eq:MSR_first_cumus}
    \left.\frac{\delta \mathcal{W}}{\delta h(t)}\right|_{\bs{j}, \bs{K}=0} = \E{x(t)}, \qquad
    \left.\frac{\delta \mathcal{W}}{\delta j(t)}\right|_{\bs{j}, \bs{K}=0} = \E{z(t)} = 0,
}
where $\E{z(t)}=0$ is a well-known property of the MSR formalism following from the normalization of the partition function \cite{hertz2016path} {(provided only a single one of the two boundary values is held fixed \cite{stapmanns2020self})}. The second cumulants follow accordingly from
\myEq{
    \left.\frac{\delta^2 \mathcal{W}}{\delta h(t)\delta j(t')}\right|_{\bs{j}, \bs{K}=0} &=  \frac{1}{c}\E{x(t)z(t')}, \\
    \left.\frac{\delta^2 \mathcal{W}}{\delta h(t)\delta h(t')}\right|_{\bs{j}, \bs{K}=0} &=  \frac{1}{c}\sbs{\E{x(t)x(t')} - \E{x(t)}\E{x(t')}}.
}
The Green function, consisting of the so-called \textit{statistical} propagator $F(t, t')$, the response functions $G^{R/A}(t, t')$ and the response-response function $Q(t, t')$, thus becomes
\begin{align*}
    \bs{G}(t, t') = 
    \begin{pmatrix}
        F(t, t') & G^R(t, t') \\
        G^A(t, t') & Q(t, t')
    \end{pmatrix}
  = \begin{pmatrix}
        \frac{\delta^2 \mathcal{W}}{\delta h(t)\delta h(t')} & \frac{\delta^2 \mathcal{W}}{\delta h(t)\delta j(t')}\\
        \frac{\delta^2 \mathcal{W}}{\delta j(t)\delta h(t')} & \frac{\delta^2 \mathcal{W}}{\delta j(t)\delta j(t')}
    \end{pmatrix},    
\end{align*}
where the {retarded} response function $G^R(t, t') = \E{x(t)z(t')} / c$ encodes the correlation of the system with {previous} noise and thus must vanish for $t'>t$. For zero sources,  the response-response propagator $Q(t, t')$ vanishes for the same reason as $\E{z(t)}$. Thus, we find that the Green function provides us with the second cumulants, i.e.\
\begin{align}\label{eq:MSR_second_cumulants}
    \left.\begin{pmatrix}
        F(t, t') & G^R(t, t') \\
        G^A(t, t') & Q(t, t')
    \end{pmatrix} \right|_{\bs{j}, \bs{K}=0}
    = \frac{1}{c}
    \begin{pmatrix}
        \E{x(t)x(t')} - \E{x(t)}\E{x(t')} & \E{x(t)z(t')} \\
        \E{z(t)x(t')} & 0
    \end{pmatrix}.
\end{align}
The special choice of fields from Eq.\ \eqref{eq:MSR_fields} also gives rise to a particular structure of the self-energy matrix. From Eq.\ \eqref{eq:2PI_self_energy}, we find that
\myEq{\label{eq:2PI_self-energy_MSR}
    \bs{\Sigma}(t, t') &= 
                        \begin{pmatrix}
                            0 & \Sigma^A(t, t') \\
                           \Sigma^R(t, t') & \Sigma^F(t, t') 
                        \end{pmatrix} 
                        = 2 c^{-1} 
                        \begin{pmatrix}
                        \frac{\delta \Gamma_2}{\delta F(t, t')} & \frac{\delta \Gamma_2}{\delta G^R(t, t')} \\
                         \frac{\delta \Gamma_2}{\delta G^A(t, t')} & \frac{\delta \Gamma_2}{\delta Q(t, t')} 
                        \end{pmatrix},
}
where the vanishing of the upper left component is again a property of the MSR formalism (also true for the Keldysh representation of the analogous quantum effective action \cite{kamenev2011field}). After expressing the diagrams contributing to $\Gamma_2$ via the propagator lines belonging to the four (scaled) cumulants $F(t, t')$, $G^{R/A}(t, t')$ and $Q(t, t')$, the equations of motion follow straightforwardly from Eqs.\ \eqref{eq:2PI_EoM}. Examples of this are given in the following.

\section{Application to Stochastic Processes}

Turning to the application of the above methods to actual physical problems, in the following we begin with a presentation of the Ornstein-Uhlenbeck process in section \ref{subsec:OU_process}, which constitutes the exactly solvable ``non-interacting'' limit of the classical field theory of stochastic processes. We then gradually increase the level of complexity by taking up the problem of geometric Brownian motion in section \ref{subsec:geom_brownian}. Even though its MSR action is formally non-linear, we find that the 2PI effective action allows us to reproduce the known exact results for the first two cumulants. Finally, we consider a scalar SDE with additive noise and a quadratic non-linearity as the natural first candidate for an interacting model in section \ref{subsec:quad_non_lin}. In place of analytical solutions, there we employ numerical integrators for SDEs \cite{DifferentialEquations.jl-2017} for comparison. 

{The general strategy is now to use Eqs.\ \eqref{eq:2PI_EoM} to derive (integro-) differential equations for the first and second cumulants. These equations will govern the time evolution of the cumulants starting from a given (Gaussian) initial condition. For the exactly solvable models, the self-consistent solutions of these equations are then equivalent to the results one would obtain from the corresponding SDE. In the non-linear case, the solutions will be correct to the specified order in the expansion parameter.}

\subsection{Ornstein-Uhlenbeck Process}\label{subsec:OU_process}

The Ornstein-Uhlenbeck process is the standard example of a stationary, Gaussian Markov process \cite{gardiner1985handbook, VanKampen2007}. It follows from the Wiener process by addition of a linear drag and is described by the SDE
\myEq{\label{eq:OU_process}
    \dd x(t) = -\lambda x(t)\dd t + \sqrt{D}\dd W(t),
}
where {$W(t > 0)$ is again a one-dimensional Brownian motion and}  $\lambda, D > 0$. Heuristically, Eq.\ \eqref{eq:OU_process} already goes back to Langevin \cite{langevin1908theorie}, where it described the \textit{velocity} of a Brownian particle in a viscous fluid under the influence of a random force. The corresponding Onsager-Machlup path integral \cite{onsager1953fluctuations} is given by
\myEq{
    \int\mathcal{D}x\exp{\cbs{-\frac{1}{2D}\mint{0}{t_f}{t}\rbs{\partial_t{x}(t) +\lambda x(t)}^2}},
}
for which Eq.\ \eqref{eq:OM_Hubbard_Straonovich} reduces to the action functional 
\myEq{\label{eq:OU_process_action}
    S[x, z] =   \mint{0}{t_f}{t}\left[ D z^2(t)/2 + \ii z(t)(\partial_t{x}(t) + \lambda x(t)) \right].
}
Note again that we employ the It\^o convention. We now obtain the inverse Green function as the Hessian of Eq.\ \eqref{eq:OU_process_action} and find
\myEq{
    \boldsymbol{G}_0^{-1}(t, t') = 
    \delta(t - t')
    \begin{pmatrix}
        0 & -\ii\partial_t + \ii\lambda \\
        \ii\partial_t + \ii\lambda & D
    \end{pmatrix}.
}
As we are discussing a Gaussian process, the classical saddle point of $S[x, z]$ provides the exact solution for the first cumulant (i.e. the mean of $x(t)$). From the saddle-point conditions
\begin{subequations}\label{eq:OU_saddle_point}
    \begin{align}
        0 &= \delta S[\bar{x}, \bar{z}]/\delta\bar{x}(t) = -\ii\partial_t \bar{z}(t) + \ii\lambda \bar{z}(t), \\
        0 &= \delta S[\bar{x}, \bar{z}]/\delta\bar{z}(t) = D\bar{z}(t) + \ii\partial_t \bar{x}(t) + \ii\lambda \bar{x}(t) \label{eq:OU_saddle_point_x}
    \end{align}
\end{subequations}
we see that a vanishing response-field average $\bar{z}(t) = 0$ is a valid solution. As mentioned below Eq.\ \eqref{eq:MSR_first_cumus}, this is a consequence of the MSR formalism. Plugging this into Eq.\ \eqref{eq:OU_saddle_point_x}, we come out with $\partial_t \bar{x}(t) = -\lambda \bar{x}(t)$ , which is also called the ``phenomenological'' law of motion \cite{onsager1953fluctuations}. 

Turning to the second cumulant or statistical propagator via Eq.\ \eqref{eq:Kadanoff-Baym}, which in the present case reduces to differential form since the self-energy vanishes, we obtain
\myEq{
    0 &= \ii\partial_t F(t, t') + \ii\lambda F(t, t') +  DG^A(t, t').
}
To find a complete solution for $F(t, t')$, one also needs the adjoint equation operating in the direction of the second time argument $t'$. Accordingly, we have the two equations
\myEq{\label{eq:OU_t_tp}
    \partial_{t\phantom{'}} F(t, t') &=  -\lambda F(t, t') + \ii DG^A(t, t'),\\
    \partial_{t'} F(t, t') &=  -\lambda F(t, t') + \ii DG^R(t, t'),
}
the second of which can be derived from the first via the symmetry properties $F(t, t') = F(t', t)$ and $G^R(t, t') = G^A(t', t)$. {The adjoint equations arise because the second cumulants are two-time functions defined on the domain $\{(t, t')\, |\, t\in[0, t_f], t'\in[0, t_f]\}$. Causality then results in different equations depending on whether $t \lessgtr t'$. The ``equal-time'' equation, for which $t=t'$, can be derived by adding Eqs.\ \eqref{eq:OU_t_tp} and then taking the limit $t'\to t$.} By transforming to so-called \textit{Wigner coordinates} $T = \myFrac{\rbs{t + t'}}{2}$, $\tau = t - t'$, Eqs.\ \eqref{eq:OU_t_tp} can also be rewritten as
\begin{subequations}\label{eq:OU_T_tau}
    \begin{align}
        \partial_{T} F(T, \tau) &=  -2\lambda F(T, \tau) + \ii D\rbs{G^A(T, \tau) + G^R(T, \tau)}, \label{eq:OU_T}\\
        \partial_{\tau} F(T, \tau) &=  \frac{\ii D}{2}\rbs{G^A(T, \tau) - G^R(T, \tau)}. \label{eq:OU_tau}
    \end{align}
\end{subequations}
The representation in Wigner coordinates is physically relevant because the stationary solutions $F^*$ will have the property of being independent of the ``center-of-mass'' time $T$, i.e. $F^*(t, t') = F^*(t - t') = F^*(\tau)$. The retarded and advanced response functions $G^{R/A}(t, t')$ are determined by
\myEq{\label{eq:OU_G_R_A}
    \delta(t - t') &= \rbs{+/-}\ii\partial_t G^{R/A}(t, t') + \ii\lambda G^{R/A}(t, t'),
}
respectively. The analytical solution for the advanced function is hence $G^A(t, t') = G^A(t - t') = -\ii\theta(t' - t)\ee^{-\lambda\rbs{t' - t}}$, which is manifestly stationary. The retarded solution follows trivially by employing the above symmetry relation. Finding the solution for $F(t, t')$ for all $t, t' > 0$ is easiest via Eqs.\ \eqref{eq:OU_T_tau}. Observing that the dynamics in $T$ decouples from that in $\tau$, we {rewrite} Eq.\ \eqref{eq:OU_T} for $\tau = 0$, i.e. 
\myEq{\label{eq:OU_equal_time}
    \partial_{T} F(T, 0) =  -2\lambda F(T, 0) +  D,
}
where we have used the identity 
\myEq{\label{eq:G_R_plus_G_A}
    G^A(T, 0) + G^R(T, 0) = G^A(t, t) + G^R(t, t) = -\ii,
} 
which is a property of the MSR formalism \cite{kamenev2011field} and thus model-independent. Now Eq.\ \eqref{eq:OU_equal_time} is solved by
\myEq{\label{eq:OU_sol_F}
    F(T, \tau) &= \rbs{F(0, 0) - \frac{D}{2\lambda}}\ee^{-2\lambda T} +  \frac{\ii D}{2\lambda}\rbs{G^A(T, \tau) + G^R(T, \tau)},
}
{where} $\tau = 0$. {However, we may employ Eq.\ \eqref{eq:OU_sol_F} also as an ansatz for the general solution at arbitrary $\tau$. To see this, observe that because} $\partial_\tau \rbs{G^A(T, \tau) + G^R(T, \tau)} = \lambda\rbs{G^A(T, \tau) - G^R(T, \tau)}$ by Eq.\ \eqref{eq:OU_G_R_A}, {the ansatz} is also the solution to Eq.\ \eqref{eq:OU_tau}. But since any point $(t, t')$ can be reached through combined time evolution via Eqs.\ \eqref{eq:OU_equal_time} and \eqref{eq:OU_tau}, we know that Eq.\ \eqref{eq:OU_sol_F} must in fact also be the solution to Eq.\ \eqref{eq:OU_T} for all $\tau$. This is confirmed upon recasting Eq.\ \eqref{eq:OU_sol_F} {into} the form
\myEq{
    F(t, t') &= F(0, 0)\ee^{-\lambda\rbs{t + t'}} - \frac{D}{2\lambda}\rbs{\ee^{-\lambda\rbs{t + t'}} - \ee^{-\lambda\myAbs{t - t'}}},
}
which is the known solution for the second cumulant of the Ornstein-Uhlenbeck process, obtained after solving Eq.\ \eqref{eq:OU_process} directly by $x(t) =  x(0)\ee^{-\lambda t} + \sqrt{D}\int_0^t \ee^{\lambda \rbs{s - t}}\dd W(s)$. To find the stationary solution, we let $T\to\infty$ and find $F^*(\tau) = D\lambda^{-1}  \ee^{-\lambda\myAbs{\tau}} / 2$. This completes our discussion of Gaussian processes.

\subsection{Geometric Brownian Motion}\label{subsec:geom_brownian}

A process that is not Gaussian yet still exactly solvable is defined by the SDE
\myEq{\label{eq:GBM_SDE}
    \dd x(t) = \mu x(t)\dd t + \sigma x(t)\dd W(t),
}
where $\mu, \sigma \in \mathds{R}$. The process $x(t)$ is called \textit{geometric Brownian motion} (GBM) and foundational for {the description of} financial markets, where it serves as the stock-price model underlying the Black-Scholes equation \cite{black1973pricing}. Beyond this, stochastic exponential growth is important in a variety of physical and biological systems, e.g. nuclear fission and population dynamics \cite{pirjol2017phenomenology}. We proceed to show that the 2PI effective action can reproduce the exact solution for the first two cumulants of this process. This can serve as a starting point for applying self-consistent perturbation theory to other non-trivial stochastic processes such as the Heston model \cite{heston1993closed}. 

The MSR action of the process $x(t)$ in It\^o convention is given by
\myEq{\label{eq:MSR_GBM}
    S[x, z] =   \mint{0}{t_f}{t}\left[ \sigma^2 x^2(t) z^2(t^+)/2 + \ii z(t)(\partial_t x(t) - \mu x(t)) \right]
}
where we have introduced time-ordering in the non-quadratic part by defining $t^\pm = t \pm \varepsilon$ for $\varepsilon \to 0^+$. {This is not strictly required because the correct regularization of the continuum limit is ensured by the causality properties of the response functions $G^{R/A}$. In the evaluation of closed time loops produced by the non-linearity, however,  some extra care is helpful to avoid ambiguities. Thus the time-ordering in Eq.\ \eqref{eq:MSR_GBM} is a reminder that the noise should always run ``ahead'' of the system \cite{kamenev2011field}.} Note that a comment about the Stratonovich convention is given below Eq.\ \eqref{eq:gbm_z_saddle}. 

\begin{figure}
	\centering
	\includegraphics[width=0.5\textwidth]{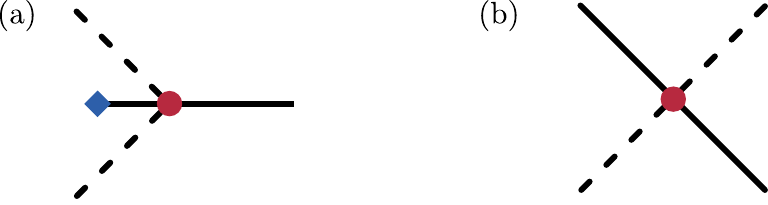}
	\caption{Graphical representation of the vertices in Eq.\ \eqref{eq:vertices}. Note that the vertex proportional to $z_0$ vanishes. Dashed lines represent the response fields $\delta z$, solid lines $\delta x$. The blue square signifies $x_0$.  (a) The effective three-vertex, (b) the four-vertex.}
\label{fig:vertices}
\end{figure}

Expanding the action around the classical saddle point, we have
\myEq{
    S[x_0 + \sqrt{c}\delta x, z_0 + \sqrt{c}\delta z] &= S[x_0, z_0] + S^{(2)} + \mathcal{V},
}
with the classical action $S[x_0, z_0]$ to lowest order. {The factors of $\sqrt{c}$ will make the Gaussian part $S^{(2)}/c$ of the exponent independent of $c$ and at the same time multiply each vertex by $c^{m/2}$, where $m$ is the number of legs. Since the expectation values in the propagator lines of the Feynman diagrams are formed with respect to $S^{(2)}/c$, this removes the need of introducing the correct prefactors of $c$ by hand. It furthermore ensures that $c$ tracks the number of loops in the loop expansion of the effective action \cite{helias2019statistical}. The effective quadratic part is then given by}
\myEq{\label{eq:gbm_delta_S}
    {S^{(2)}/c} = \mint{0}{t_f}{t}&\left[\sigma^2 x_0^2\delta z^2(t)/2 + \ii\delta z(t)\rbs{\partial_t - \mu}\delta x(t) \right. \\
     &+ \left. 2\sigma^2 z_0 x_0\delta z(t^+)\delta x(t) + \sigma^2 z_0^2 \delta x^2(t) {/} 2 \right],
}
and the non-quadratic part
\myEq{\label{eq:vertices}
     {\mathcal{V}/c} &= \sigma^2\mint{0}{t_f}{t}\left[\sqrt{c}\rbs{z_0\delta z(t^+)\delta x^2(t) + x_0\delta x(t)\delta z^2(t^+)} + c\delta x^2(t)\delta z^2(t^+)/2\right],
}
where the vertex given by the first term does not contribute to the perturbation series because it is proportional to $z_0$. {Observe that terms proportional to powers of $z_0$ may always be discarded immediately, \textit{provided} one does not have to differentiate with respect to them at a later point.} A graphical representation of the two remaining cubic and quartic vertices is provided in Fig.\ \ref{fig:vertices}. 
The inverse Green function is given by the Hessian of $S^{(2)} / c$ in Eq.\ \eqref{eq:gbm_delta_S}, evaluated at the Legendre-transformed fields, i.e.
\myEq{\label{eq:gbm_G_0_inv}
    \boldsymbol{G}_0^{-1}[\bar{x}, \bar{z}] (t, t')&= 
    \delta(t - t')
    \begin{pmatrix}
        \sigma^2\bar{z}^2 & -\ii\partial_t - \ii\mu \\
        \ii\partial_t - \ii\mu & \sigma^2\bar{x}^2
    \end{pmatrix}
    \\
    &+
   2\sigma^2 \bar{z} \bar{x}\begin{pmatrix}
        0 & \delta(t^+ - t') \\
        \delta(t^- - t') & 0
    \end{pmatrix},
}
where in the second term we have taken care of the time-ordered fields $\delta z(t^+) \delta x(t) = \delta z(t) \delta x(t^-)$ in Eq.\ \eqref{eq:gbm_delta_S}. 

\begin{figure}
	\centering
	\includegraphics[width=0.6\textwidth]{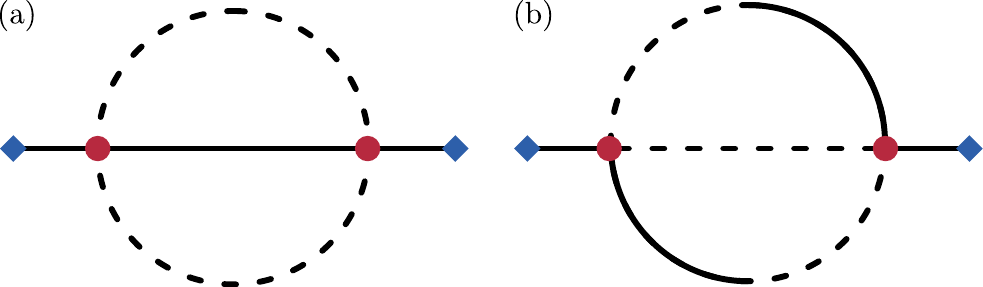}
	\caption{Second-order diagrams arising from the cubic vertex of Fig.\ \ref{fig:vertices}(a). (a) The self-energy following by differentiation from this graph necessarily vanishes because it contains at least one response-response propagator. (b) In this case, the self-energy must also vanish, either because of one response-response propagator or because of a retarded-advanced loop.}
\label{fig:three_vertex_second_order}
\end{figure}

To find the 2PI diagrams contributing to $\Gamma_2$ in Eq.\ \eqref{eq:2PIEA}, we first consider the cubic vertex depicted in Fig.\ \ref{fig:vertices}(a). {For a given number of vertices} $2k$, $k\in\mathds{N}$, this vertex has $4k$ response fields and $2k$ fields $\delta x(t)$. Thus for $k\geq 2$, there are at least two response-response propagators $Q(s, s')$ contributing to any connected diagram, where $s,s'$ are internal times. According to Eqs.\ \eqref{eq:MSR_second_cumulants} and \eqref{eq:2PI_self-energy_MSR}, after differentiation any of the self-energy components $\Sigma^{R/A/F}$ will hence {contain} at least one vanishing response-response propagator. For $k=1$ we find the two diagrams shown in Fig.\ \ref{fig:three_vertex_second_order}.
The self-energies derivable from Fig.\ \ref{fig:three_vertex_second_order}(a) vanish by the same argument we have just given for the case $k\geq 2$, while Fig.\ \ref{fig:three_vertex_second_order}(b) induces either a vanishing response-response propagator or a closed retarded-advance loop of the form $G^R(s, s')G^A(s, s')$, which must also vanish due to causality. To summarize, we have shown that the cubic vertex of Fig.\ \ref{fig:vertices}(a) does not yield any non-trivial self-energies.
\begin{figure}
	\centering
	\includegraphics[width=0.6\textwidth]{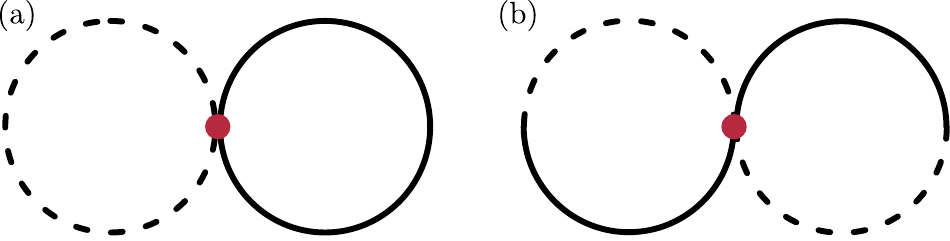}
	\caption{Diagrams corresponding to $\Gamma_2$ as given in Eq.\ \eqref{eq:four_vertex_first_order}. (a) This diagram leads to the self-consistent Hartree-Fock approximation for the self-energy $\Sigma^F$. (b) As this diagram consists of two equal-time loops involving retarded (advanced) propagators, it vanishes due to time-ordering.}
\label{fig:four_vertex_first_order}
\end{figure}

\begin{figure}
	\centering
	\includegraphics[width=0.65\textwidth]{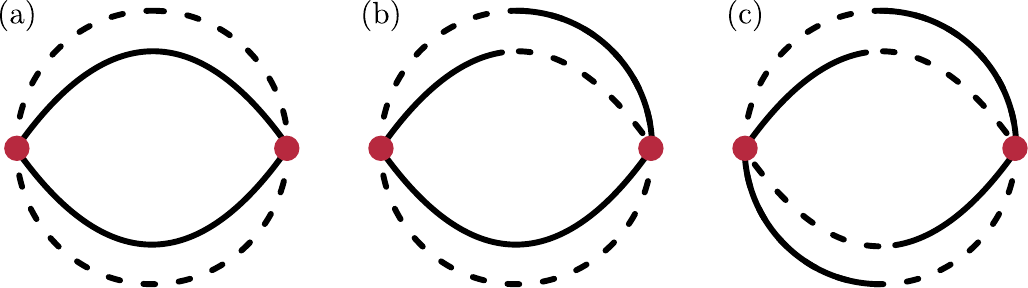}
	\caption{Connected, 2PI second-order diagrams arising from the  quartic vertex of Fig.\ \ref{fig:vertices}(b). Any self-energy derivable from these diagrams must vanish either on account of a response-response propagator or a closed retarded-advanced loop.}
\label{fig:four_vertex_second_order}
\end{figure}

Turning to the quartic vertex in Fig.\ \ref{fig:vertices}(b), to lowest order we find the two diagrams shown in Fig.\ \ref{fig:four_vertex_first_order}, which correspond to the self-consistent Hartree-Fock approximation in the framework of the 2PI effective action. Written explicitly in terms of propagators, these diagrams are given in Eq.\ \eqref{eq:four_vertex_first_order}. To second-order, the quartic vertex gives rise to the diagrams illustrated in Fig.\ \ref{fig:four_vertex_second_order}. Again, because of response-response propagators and closed retarded-advanced loops, these diagrams do not result in non-trivial self-energies. 

This leaves us to investigate any connected, 2PI diagram with $l > 2$ quartic vertices, where $l\in \mathds{N}$, as well as all possible combinations of cubic and quartic vertices. The latter diagrams are formed out of $2k$ cubic and $l$ quartic vertices, which as before rules out all diagrams with $k\geq 2$. We do not attempt a formal proof that all of these diagrams vanish since this is achieved more directly via the recovery of the exact solution for GBM from the diagrams in Fig.\ \ref{fig:four_vertex_first_order}.

To still gain some insight into why these diagrams vanish, consider Fig.\ \ref{fig:four_vertex_second_order}(a). If we break up one solid and one dashed propagator line, we can insert another quartic vertex from Fig.\ \ref{fig:vertices}(b) and connect solid to solid and dashed to dashed legs. Topologically, this produces the only connected, 2PI diagram at $l=3$. All other valid diagrams with three quartic vertices follow from the latter by opening and mixing any pair of propagators (in the same way Fig.\ \ref{fig:four_vertex_second_order}(b) can be obtained from Fig.\ \ref{fig:four_vertex_second_order}(a) by opening two propagator lines and ``twisting'' them.) Clearly, any of the self-energies derivable from the diagrams with $l=3$ must vanish, analogously to those of Fig.\ \ref{fig:four_vertex_second_order}. 

The simplest diagrams containing both cubic and quartic vertices are produced by cutting the two propagators in the first-order diagrams of Fig.\ \ref{fig:four_vertex_first_order} and connecting the resulting amputated legs in all ways to those of two cubic vertices joined by a single propagator. These diagrams are connected, 2PI, and third order in the coupling $\sigma^2$. As one may verify, all of them contain at least one response-response propagator and a retarded-advanced loop, such that the resulting self-energies again vanish.

Summing up, we assume that the only diagrams contributing non-trivially to the self-energy matrix are those of Fig.\ \ref{fig:four_vertex_first_order}, and consequently that the \textit{exact} 2PI effective action of GBM {in It\^o convention} is found {by} letting
\myEq{\label{eq:four_vertex_first_order}
    \Gamma_2 = \frac{c^2\sigma^2}{2}\mint{}{}{t} Q(t, t) F(t, t) + \frac{c^2\sigma^2}{4}\mint{}{}{t}\rbs{G^A(t^+, t) + G^R(t^-, t)}^2,
}
where we keep the second {term for} completeness (because of causality it does not contribute to the self-energies {in the It\^o regularization where $G^A(t^+, t) = G^R(t^-, t) = 0$}). {To go from the vertices on the right-hand side of Eq.\ \eqref{eq:vertices} to the functional in Eq.\ \eqref{eq:four_vertex_first_order}, the diagrammatic rules were as follows: 1) Take the power of each vertex at the given order in the expansion and contract the fields in all possible combinations according to Wick's theorem \cite{helias2019statistical}, i.e.\ $\delta x^2(t)\delta z^2(t^+) \to \{\E{\delta x^2(t)}\E{\delta z^2(t^+)}, 2\E{\delta x(t)\delta z(t^+)}^2\}$. 2) Replace the averages by Green functions: $\{Q(t, t) F(t, t), [G^A(t^+, t) + G^R(t^-, t)]^2/2\}$, where we also have symmetrized the response functions. 3) Multiply by the {prefactors determined by the order of the interaction}, in this case $c\cdot  c\sigma^2/2$.} 

To derive the phenomenological law of motion \cite{onsager1953fluctuations} for the process $x(t)$, we now take the derivative of $\Gamma[\bs{\bar{\phi}}, \bs{ G}]$ according to Eq.\ \eqref{eq:2PI_EoM} with $\bs{\bar{\phi}} = (\bar{x}, \bar{z})^T$, i.e.
\myEq{\label{eq:gbm_x_saddle}
    0 &= \frac{\delta S[\bar{x}, \bar{z}]}{\delta \bar{z}(t)} + \frac{c}{2}\int\dd t^{\prime}\dd t^{\prime\prime}\,\operatorname{Tr}\frac{\delta \boldsymbol{G}_0^{-1}(t^{\prime}, t^{\prime\prime})}{\delta \bar{z}(t)}\boldsymbol{ G}(t^{\prime\prime}, t^{\prime}) \\
    &= \sigma^2 \bar{z} \bar{x}^2 + \rbs{\ii\partial_t -\ii\mu}\bar{x} + c\sigma^2 \bar{x}\sbs{G^A(t^+, t) + G^R(t^-, t)}.
}
The time-ordering in the response functions ensures that they evaluate to zero, such that with $\bar{z}(t)=0$ we find $\partial_t\bar{x}(t)=\mu\bar{x}(t)$, as expected for GBM in It\^o convention. That $\bar{z}(t)=0$ is indeed a solution can be verified by calculating
\myEq{\label{eq:gbm_z_saddle}
    0 &= \frac{\delta S[\bar{x}, \bar{z}]}{\delta \bar{x}(t)} + \frac{c}{2}\int\dd t^{\prime}\dd t^{\prime\prime}\,\operatorname{Tr}\frac{\delta \boldsymbol{G}_0^{-1}(t^{\prime}, t^{\prime\prime})}{\delta \bar{x}(t)}\boldsymbol{ G}(t^{\prime\prime}, t^{\prime}) \\
    &= \sigma^2 \bar{x} \bar{z}^2 - \rbs{\ii\partial_t  + \ii\mu}\bar{z} + c\sigma^2 \bar{z}\sbs{G^A(t^+, t) + G^R(t^-, t)},
}
where the contribution from the first term in Eq.\ \eqref{eq:gbm_G_0_inv} vanishes because it is traceless. {Note that using the Stratonovich convention would lead to a different first cumulant since we are studying an SDE with \textit{multiplicative} noise. Specifically, one expects a solution $\bar{x}(t) = \bar{x}(0)\exp{\rbs{(\mu + \sigma^2/2) t}}$, i.e.\ $\partial_t\bar{x}(t)=(\mu + \sigma^2/2)\bar{x}(t)$. This would result from an additional noise-dependent drift term $\sigma^2 x(t) / 2$ in the Stratonovich analogue of Eq.\ \eqref{eq:MSR_GBM}, and a modified regularization of the response functions, i.e.\ $G^{R/A}(t^{-/+}, t) = G^{R/A}(t, t) = -i/2$ in our notation {(s.\ also Ref.\ \cite{hertz2016path} and section 4.4.2 of Ref.\ \cite{gardiner1985handbook})}.} 

To find the self-energy, we plug $\Gamma_2$ from Eq.\ \eqref{eq:four_vertex_first_order} into Eq.\ \eqref{eq:2PI_self-energy_MSR} to obtain
\myEq{\label{eq:2PI_self-energy_GBM}
    \bs{\Sigma}(t, t') &=  -c \sigma^2 \delta(t - t')
    \begin{pmatrix}
        0 & { G^R(t^-, t)} \\
        { G^A(t^+, t)} & F(t, t)
    \end{pmatrix}.
}
Taking into account that $G^R(t^-, t) = G^A(t^+, t) = 0$, the response self-energies $\Sigma^{R/A}(t, t')$ vanish, while the lower-left component of Eq.\ \eqref{eq:Kadanoff-Baym} leads to
\myEq{
    0 &= \rbs{\ii\partial_t - \ii\mu} F(t, t') + \sigma^2\bar{x}(t)^2 G^A(t, t')  - \mint{0}{t_f}{s}\Sigma^F(t, s)G^A(s, t').
}
Writing down also the adjoint {(cf.\ Eqs.\ \eqref{eq:OU_t_tp})}, we thus have the time-local equations
\begin{subequations}\label{eq:gbm_eqm_F}
    \begin{align}
        \partial_t F(t, t') &= \mu F(t, t') + \ii \sigma^2\sbs{ \bar{x}^2(t)  + cF(t, t)}G^A(t, t'), \label{eq:gbm_eqm_F_t} \\
        \partial_{t'} F(t, t') &= \mu F(t, t') + \ii \sigma^2\sbs{\bar{x}^2(t') + cF(t', t')}G^R(t, t').
    \end{align}
\end{subequations}

Since the scaling constant $c$ was introduced for formal reasons, we may now set $c = 1$ and transform Eqs.\ \eqref{eq:gbm_eqm_F} to Wigner coordinates. In the direction of the center-of-mass time $T$, this leaves us with
\myEq{
    \partial_T F(T, \tau) &= 2\mu F(T, \tau) 
    + \ii \sigma^2 F(T, 0)\sbs{G^A(T, \tau) + G^R(T, \tau)} \\
    &+ \ii \sigma^2\sbs{\bar{x}^2(T + \tau/2)G^A(T, \tau) + \bar{x}^2(T-\tau/2)G^R(T, \tau)}.
}
In the equal-time limit $\tau\to 0$, and with the identity in Eq.\ \eqref{eq:G_R_plus_G_A}, this transforms into
\myEq{
    \partial_t F(t, t) &= 2\mu F(t, t) + \sigma^2\sbs{ \bar{x}^2(t)  + F(t, t)},
}
which is solved by $F(t, t) = \bar{x}^2(t)\sbs{\exp{\rbs{\sigma^2 t}} - 1}$ where $\bar{x}(t) = \bar{x}(0)\exp{\rbs{\mu t}}$. For $t>t'$,  Eq.\ \eqref{eq:gbm_eqm_F_t} has the solution $F^>(t - t', t') = F_0\exp{\rbs{\mu \rbs{t -t'}}} $. With an initial condition $F_0 = F(t', t')$ placed on the equal-time diagonal, we can thus find the general solution
\myEq{\label{eq:gbm_sol_F}
    F(t, t') &= \bar{x}(t)\bar{x}(t')\sbs{\exp{\rbs{\sigma^2 t'}} - 1},
}
which holds for $t\geq t'$. The solution on the upper triangle of the two-time square $(t, t')$ can again be found from symmetry. The general solution of the It\^o SDE \eqref{eq:GBM_SDE} is {(e.g.\ section 4.4.7 of Ref.\ \cite{gardiner1985handbook})}
\myEq{
    x(t) &= x(0)\exp\rbs{(\mu - \sigma^2/2)t}\exp\rbs{\sigma(W_t - W_0)}.
}
Since the variance of $(W_t - W_0) + (W_{t'} - W_0) = \rbs{W_t - W_{t'}} + 2\rbs{W_{t'} - W_0}$ is given by $\E{\rbs{W_t - W_{t'}}^2} + 4\E{\rbs{W_{t'} - W_0}^2} = t - t' + 4t'$ for $t\geq t'$, one has 
{\myEq{
    \E{x(t)x(t')} &= \bar{x}^2(0)\exp\rbs{(\mu - \sigma^2/2)(t+t')}\left\langle{\exp\rbs{\sigma(W_t - W_0 + W_{t'} - W_0)}}\right\rangle \\
    &= \bar{x}(t)\bar{x}(t')\exp\rbs{\sigma^2 \sbs{t + 3t' - (t+t')} / 2} = \bar{x}(t)\bar{x}(t')\exp\rbs{\sigma^2 t'},
}}
and consequently Eq.\ \eqref{eq:gbm_sol_F} as the solution for the two-time variance of the process $x(t)$. As mentioned above, this also serves as proof that in the MSR formalism, the entire diagrammatic series of GBM vanishes (apart from the single diagram in Fig.\ \ref{fig:four_vertex_first_order}(a)).

\subsection{Quadratic Non-Linearity with Additive Noise}\label{subsec:quad_non_lin}

\begin{figure}
	\centering
	\includegraphics[width=0.6\textwidth]{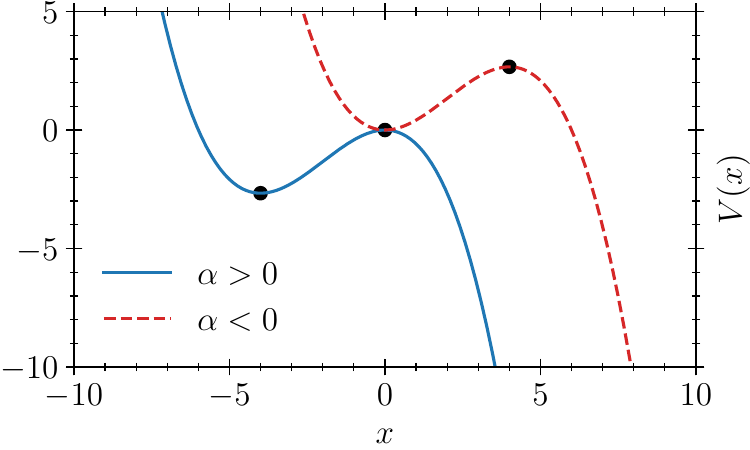}
    \caption{The classical potential $V(x) = -(\alpha x^2 / 2 + \beta x^3 / 3)$ corresponding to Eq.\ \eqref{eq:sde_cubic} for $\alpha \lessgtr 0$. Observe that for $D=0$, the trajectories seek to minimize $V(x)$ and stop at positions for which $V'(x)=0$ (cf. Ref.\ \cite{gardiner1985handbook}). The numerical values for the parameters in the plot are $\myAbs{\alpha} = 1$, $\beta=1/4$.}
    \label{fig:quad_non_lin_potential}
\end{figure}

\begin{figure}[!t]
	\centering
	\includegraphics[width=0.6\textwidth]{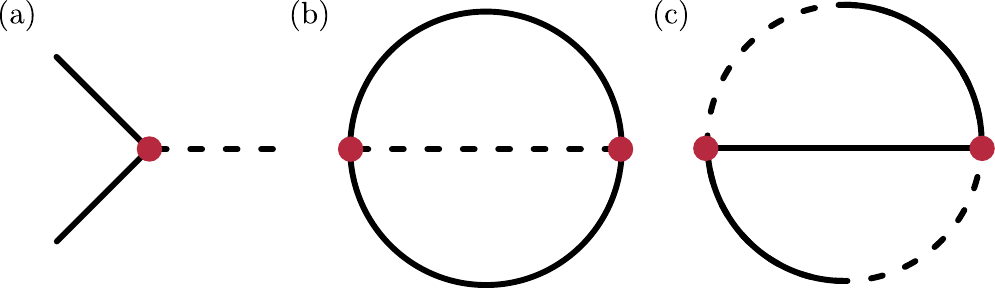}
    \caption{(a) The three-vertex of the MSR action defined in Eq.\ \eqref{eq:quad_non_lin_action}. (b, c) Second-order diagrams contributing to the 2PI effective action. The ``dumbbell'' graphs \cite{carrington20044pi}, which can be obtained from (b, c) by opening two lines and forming two closed loops connected by a single propagator, are not 2PI, for instance, and hence not part of $\Gamma_2$.}
    \label{fig:quad_non_lin_diagrams}
\end{figure}

A natural starting point for the investigation of the performance of the 2PI effective action on a non-linear model is provided by the SDE
\myEq{\label{eq:sde_cubic}
    \dd x(t) &= \rbs{\alpha x(t) + \beta x^2(t)} \dd t + \sqrt{D}\dd W(t).
}
Note that this choice furthermore allows for a direct comparison to previous work involving the 1PI effective action \cite{stapmanns2020self}. The corresponding classical potential $V(x) = -(\alpha x^2 / 2 + \beta x^3 / 3)$ is plotted in Fig.\ \ref{fig:quad_non_lin_potential} for both positive and negative values of $\alpha$. The fixed points of this map are located at $x^*_{1} = 0$ and $x^*_{2} = -\alpha/\beta$. Stability is given whenever $\alpha + 2\beta x < 0$, which is true for $x < 0$ at $\alpha > 0$ and $x<\myAbs{\alpha}/\beta$ at $\alpha < 0$. It is thus simplest to set $\alpha > 0$ and confine the stable region to negative positions. Physically, the escape of a trajectory to infinity can serve, among other things, as a model of a firing neuron \cite{stapmanns2020self}. The MSR action of Eq.\ \eqref{eq:sde_cubic} in It\^o form is then given by
\myEq{\label{eq:quad_non_lin_action}
    S[x, z] &= \mint{0}{t_f}{t}\sbs{\myFrac{Dz(t)^2}{2} + \ii z(t) \rbs{\partial_t x(t) - \alpha x(t) - \beta x(t^-)^2}},
}
where we have again applied the appropriate time-ordering in the non-linear part. The graphical representation of the interaction vertex is shown in Fig.\ \ref{fig:quad_non_lin_diagrams}(a). 

One may now ask in which small parameter the expansion is to be performed, i.e. which parameter is to play the role of the so-far purely formal constant $c$? The answer is found by rewriting Eq.\ \eqref{eq:quad_non_lin_action} in the following way:
\begin{align*}
    S[x, z] &= \frac{\alpha}{\beta}\mint{0}{(\beta/\alpha)\; \alpha t_f}{s}\sbs{\myFrac{(D/\alpha)z(s)^2}{2} + \ii z(s) \rbs{(\beta/\alpha)\partial_s x(s) -x(s) - (\beta/\alpha) x(s^-)^2}},    
\end{align*}
where $s$ is a now unitless parameter. Thus we see that $\beta/\alpha$ can play the role of a small number dividing the exponent, under the condition that all frequencies are measured in units of the drift $\alpha$, and that both the integration interval and the time derivative are rescaled by $\beta/\alpha$. Practically, of course, we may simply work with Eq.\ \eqref{eq:quad_non_lin_action} and set $c=1$, which we do in the following {(this does \textit{not} imply that $\beta/\alpha = 1$)}. Proceeding as before, we find an inverse Green function
\myEq{\label{eq:quad_non_lin_G_0_inv}
    \boldsymbol{G}_0^{-1}(t, t') &= 
    \delta(t - t')
    \begin{pmatrix}
        -2\ii\beta \bar{z} & -\ii\partial_t - \ii\alpha \\
        \ii\partial_t - \ii\alpha & D
    \end{pmatrix}
    \\
    &-
   2\ii\beta\bar{x}\begin{pmatrix}
        0 & \delta(t^+ - t') \\
        \delta(t^- - t') & 0
    \end{pmatrix}.
}
Analogously to Eq.\ \eqref{eq:gbm_z_saddle}, this does not yield a contribution to the equation of motion for $\bar{z}(t)$. Instead of Eq.\ \eqref{eq:gbm_x_saddle}, we now find
\myEq{\label{eq:quad_non_lin_mf_eq}
    0 &= \partial_t \bar{x} - \alpha \bar{x} - \beta \bar{x}^2 - \beta F(t, t).
}
Note that this equation contains the \textit{resummed} statistical propagator $F$ instead of the bare one. It thus results in a self-consistently corrected steady-state $\bar{x}^*$, specified by the quadratic equation $0 = \alpha\bar{x}^* + \beta\rbs{ \bar{x}^*}^2 + \beta F^*$, the solution to which is
\myEq{\label{eq:quad_non_lin_x_saddle}
    \bar{x}^* = -\frac{\alpha}{2\beta} \pm \frac{1}{2\beta}\sqrt{\alpha^2 - 4\beta^2 F^*}.
}
Hence, while for the practically viable one-loop approximation of the 1PI effective action employed in \cite{stapmanns2020self}, the influence of the fluctuations on the steady-state of the mean is provided by the \textit{bare} correlation function, the 2PI effective action takes the corresponding corrections to the mean into account automatically when second-order diagrams are included.

To \textit{first order} in the interaction parameter $\beta$, the statistical propagator obeys the equation of motion
\myEq{\label{eq:quad_non_lin_stat_GF_first_order}
    0 &= \rbs{\partial_{t\phantom{'}} -\alpha - 2\beta \bar{x}}F(t, t') - \ii D G^A(t, t'), 
}
which in the equal-time limit gives
\myEq{
    \partial_t F(t, t) &= \rbs{2\alpha + 4\beta \bar{x}}F(t, t) + D.
}
Hence, we find the steady-state solution $F^* = -\myFrac{D}{\rbs{2\alpha + 4\beta\bar{x}^*}}$. Inserting this into Eq.\ \eqref{eq:quad_non_lin_x_saddle} reproduces the result from Ref.\ \cite{stapmanns2020self}. Note also that since $F^* > 0$, the parameters $\alpha$ and $\beta$ need to be chosen such that the denominator of $F^*$ is always negative, which coincides with the stability condition found earlier. {For completeness, we note that to this order, the retarded} Green function is determined by
{
\myEq{\label{eq:quad_non_lin_advanced_GF}
    -\ii \delta(t - t') = \rbs{\partial_{t} -\alpha - 2\beta \bar{x}} G^R(t, t').
}
}An approximation beyond mean-field plus quadratic fluctuations is obtained {upon} including the \textit{second-order} diagrams shown in Fig.\ \ref{fig:quad_non_lin_diagrams}(b, c). Writing down the corresponding contributions to the 2PI effective action, we find
\myEq{\label{eq:quad_non_lin_EA_second_order}
    \Gamma_2 &= \beta^2\int\dd t\dd t'\;\sbs{2G^R(t, t') G^A(t, t')F(t, t') + F^2(t, t') Q(t, t')} + \mathcal{O}(\beta^4).
}
Again taking derivatives with respect to the Green functions, we obtain the one-loop self-energy
\myEq{
    \bs{\Sigma}(t, t') &= 
                        - \beta^2
                        \begin{pmatrix}
                            0 & 4 G^A(t, t') \\
                            4 G^R(t, t') & 2 F(t, t')
                        \end{pmatrix} F(t, t').
}
{Eq.\ \eqref{eq:quad_non_lin_advanced_GF} now turns into an integro-differential equation
\myEq{
    -\ii \delta(t - t') = \rbs{\partial_{t} -\alpha - 2\beta \bar{x}} G^R(t, t') - 4\ii\beta^2 \mint{0}{t}{s} F(t, s) G^R(t, s) G^R(s, t'),
}
}{while Eq.\ \eqref{eq:quad_non_lin_stat_GF_first_order} becomes accordingly}
\myEq{\label{eq:quad_non_lin_stat_GF_second_order}
    0 &= \rbs{\partial_{t\phantom{'}} -\alpha - 2\beta \bar{x}}F(t, t') - \ii D G^A(t, t') \\
    &-\ii\beta^2 \mint{0}{t}{s}\cbs{4 G^R(t, s) F(t,s) F(s, t') + 2 F^2(t, s)G^A(s, t')}.
}
Observe that the second term under the integral only needs to be evaluated up to $t'$ (from causality of $G^A$, and assuming $t\geq t'$). In the equal-time limit, Eq.\ \eqref{eq:quad_non_lin_stat_GF_second_order} leads to
\myEq{\label{eq:quad_non_lin_stat_GF_second_order_tt}
    \partial_t F(t, t) &= \rbs{2\alpha + 4\beta \bar{x}}F(t, t) + D + 12 \beta^2 \mint{0}{t}{s} \mathcal{K}(t, s) F^2(t, s),
}
where the symmetric integral kernel $\mathcal{K}(t, t')$ obeys the equation
{
\myEq{\label{eq:quad_non_lin_kernel}
    0 &= \rbs{\partial_t - \alpha - 2\beta\bar{x}}\mathcal{K}(t, t') - 4\beta^2 \mint{t'}{t}{s} F(t, s) \mathcal{K}(t, s) \mathcal{K}(s, t')
}
}for $t \geq t'$, such that $G^R(t, t') = -\ii \theta(t - t') \mathcal{K}(t, t')$ and $G^A(t, t') = -\ii \theta(t' - t)\mathcal{K}(t', t)$, where $\mathcal{K}(t', t) = \mathcal{K}(t, t')$. {Note that the integral term in Eq.\ \eqref{eq:quad_non_lin_kernel} is expected to be small because the overlap of the two kernels tends to be negligible.} This kernel may also be employed in the solution of Eq.\ \eqref{eq:quad_non_lin_stat_GF_second_order} since 
\myEq{
\mint{0}{t}{s} \ii G^R(t, s) = \mint{0}{t}{s} \mathcal{K}(t, s), \qquad \mint{0}{t}{s} \ii G^A(s, t') = \mint{0}{t'}{s} \mathcal{K}(s, t'),
} 
again assuming w.l.o.g.\ that $t\geq t'$. On the equal-time diagonal, the kernel is constant and unity, i.e. $\mathcal{K}(t, t) = 1$. 

The coupled set of non-linear Eqs.\ \eqref{eq:quad_non_lin_mf_eq}, \eqref{eq:quad_non_lin_stat_GF_second_order}, \eqref{eq:quad_non_lin_stat_GF_second_order_tt} and \eqref{eq:quad_non_lin_kernel} needs to be solved self-consistently. This can be achieved efficiently and accurately by employing, for instance, a multi-step predictor-corrector method on a fixed-step two-dimensional grid \cite{bock2016buildup, lappe2021non}. Further discussions of numerical methods for equations of this type (in many-body physics usually referred to as \textit{Kadanoff-Baym} equations \cite{Baym1961}) can be found in Refs.\ \cite{dahlen2007solving, Stan2009, balzer2011solving, meirinhos2021adaptive}.

As a comparison for our numerical solutions of the above integro-differential equations, we employ state-of-the-art Runge-Kutta methods for SDEs \cite{rossler2010runge, rossler2010strong}. Specifically, we use implementations \cite{DifferentialEquations.jl-2017, bezanson2017julia} of the algorithms SRI1W1 and SRI2W1 derived in Ref.\ \cite{rossler2010runge}. Both algorithms converge with strong order 1.5, while their weak orders are 2.0 and 3.0, respectively. {Roughly speaking, in the strong case, one is concerned with the convergence of an approximate stochastic process towards the true process, while weak convergence addresses the convergence of expectation values of functionals of such processes \cite{kloeden2013convergence}.}

As an analytical comparison, we furthermore use the stationary solutions for the mean and the variance derived in one-loop approximation from the 1PI effective action in Ref.\ \cite{stapmanns2020self}, which we denote by $\bar{x}^*_{\text{1PI}}$ and $F^*_{\text{1PI}}$, respectively. Note that the 1PI one-loop approximation for the propagators is also of \textit{second} order in $\beta$.

\begin{figure}[!htb]
	\centering
	\includegraphics[width=0.9\textwidth]{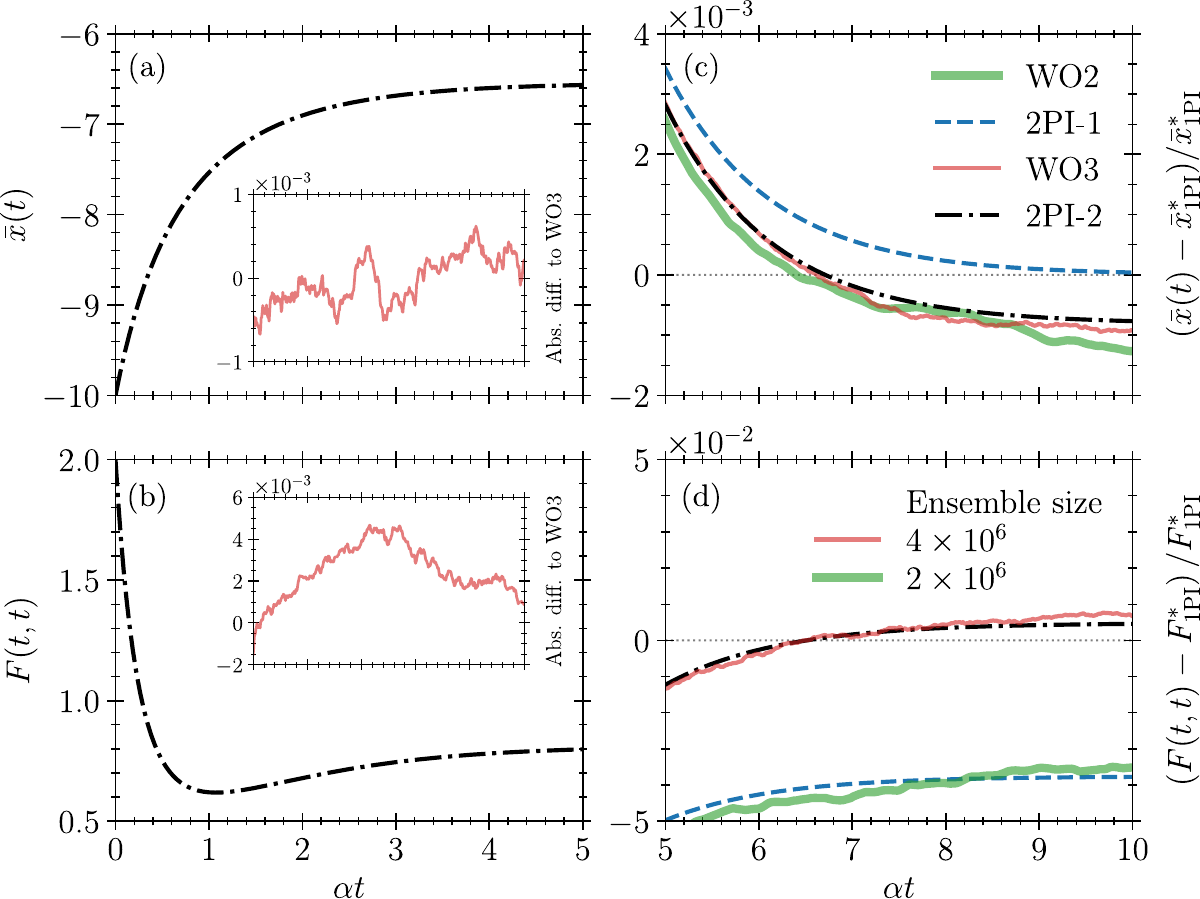}
    \caption{Comparison of the numerical solutions to Eqs.\ \eqref{eq:quad_non_lin_mf_eq}, \eqref{eq:quad_non_lin_stat_GF_second_order}, \eqref{eq:quad_non_lin_stat_GF_second_order_tt} and \eqref{eq:quad_non_lin_kernel} with the statistics obtained from a stochastic Runge-Kutta simulation of Eq.\ \eqref{eq:sde_cubic} for $D/\alpha = 1.5$, $\beta/\alpha = 0.15$, $\bar{x}(0) = -10$ and $F(0, 0) = 2$. (a, b) Time evolution of the first two cumulants and (insets) absolute differences to the algorithm SRI2W1. (c) Mean $\bar{x}(t)$ obtained via different methods, where 2PI-1 and -2 denote the first- and second-order approximations to the 2PI effective action. WO2 and WO3 denote the algorithms SRI1W1 and SRI2W1 \cite{rossler2010runge}, which converge with weak order 2.0 and 3.0, respectively. (d) Equal-time variance $F(t, t)$ from the same methods as in (c). The ensemble sizes for the SDE algorithms are indicated in (d).}
    \label{fig:quad_non_lin_dynamics}
\end{figure}

Exemplary results are presented in Fig.\ \ref{fig:quad_non_lin_dynamics} for parameters $D/\alpha = 1.5$ and $\beta/\alpha = 0.15$. The integro-differential equations were solved on a two-dimensional grid with fixed step size $\alpha\Delta t = 2^{-10}$, while the SDE solvers implemented in Ref.\ \cite{DifferentialEquations.jl-2017} work with adaptive step sizes. {Our step size was chosen small to ensure convergence given that the integrals in the equations of motion are evaluated with the trapezoidal rule only. However, the steady-states shown in Fig.\ \ref{fig:quad_non_lin_dynamics} in fact converge reasonably for a step size of $\alpha\Delta t = 2^{-8}$. Advanced numerical methods for integro-differential equations, employing higher-order integrators and thus enabling larger step sizes, are discussed in Ref.\ \cite{meirinhos2021adaptive}.} 

For the SDE solvers SRI1W1 and SRI2W1, we fixed the ensemble size to $2es 10^6$ and $4es 10^6$, respectively. Divergent trajectories escaping to $x\to\infty$ were discarded and replaced. In this way, in order to reach two (four) million trajectories in total, 57 (102) divergent trajectories were discarded. Since these numbers correspond to negligible fractions, this is equivalent to only considering non-escaping trajectories \cite{stapmanns2020self}. {Note that we used a greater ensemble size in the comparison with SRI2W1 because it serves to confirm one of our main results.}

Figs.\ \ref{fig:quad_non_lin_dynamics}(a, b) show our results for the mean and variance from the second-order 2PI effective action (2PI-2), starting from initial conditions $\bar{x}(0) = -10$ and $F(0, 0) = 2$. The insets show the absolute difference to the statistics obtained via the SDE solver with weak order 3.0 (WO3), thus confirming the overall accuracy of the solutions obtained from Eqs.\ \eqref{eq:quad_non_lin_mf_eq}, \eqref{eq:quad_non_lin_stat_GF_second_order}, \eqref{eq:quad_non_lin_stat_GF_second_order_tt} and \eqref{eq:quad_non_lin_kernel}. Fig.\ \ref{fig:quad_non_lin_dynamics}(c) summarizes the results for the mean $\bar{x}(t)$, where we also show how the first-order approximation of the 2PI effective action (2PI-1) recovers $\bar{x}^*_{\text{1PI}}$ in the long-time limit. Here, both of the SDE solvers reproduce our solution 2PI-2. This is not true once we move to Fig.\ \ref{fig:quad_non_lin_dynamics}(d), where only the algorithm of weak order 3.0 (SRI2W1) captures the second-order result from the 2PI effective action. The algorithm of weak order 2.0 (SRI1W1) instead reproduces the \textit{first-order} approximation 2PI-1. In comparison to the one-loop 1PI result $F^*_{\text{1PI}}$, we see that the additional diagrams resummed by the 2PI effective action lead to a systematic improvement of about half a percent, which is expected to become more pronounced at larger non-linearity. {The different effective actions are thus found to converge to different stationary statistics at large final times.}

\section{Conclusion}\label{sec:conclusion}

We have developed the 2PI effective action into a form such that it can be applied to non-linear stochastic processes. This allows one to derive self-consistent equations of motion for their first and second cumulants. Instead of calculating Feynman diagrams analytically in terms of the second cumulants of a Gaussian process, {as is done in the 1PI approach,} the self-consistency for the second {cumulants} instead leads to non-linear integro-differential equations. On one hand, while the perturbative expansion is simplified in this way, an analytical solution of these equations is in general not possible.  Since the equations are causal and thus depend only on the history of the process, they can, on the other hand, be solved straightforwardly by explicit time-stepping schemes. Accordingly, we found that the 2PI effective action provides accurate results for moderate non-linearities at low computational costs. Compared to the 1PI results for the stationary state of the system, we saw that the latter performs well, even though the additional diagrams summed up by the 2PI effective action provided additional accuracy.

{As the extended Plefka transformation of Ref.\ \cite{Bravi_2016} is related to the present work through the use of a second Legendre transformation, it is useful to discuss some differences to our work. The most important one is that the presentation in Ref.\ \cite{Bravi_2016} is limited to an expansion of second order in a parameter quantifying the interaction strength. Even though the same is true coincidentally for our discussion in section \ref{subsec:quad_non_lin}, there is, in principle, no added difficulty (except numerically) in expanding the self-energy to higher orders, as has indeed been done in QFT applications \cite{Schl_nzen_2019}. Also, (non-perturbative) expansions in other parameters than the interaction strength, such as $1/\mathcal{N}$ expansions \cite{Berges2004, Berges2015}, follow directly within our framework. While higher-order expansions, in particular, should also be possible via the extended Plefka transformation, it seems that this would be more complicated without the notions of effective action and self-energy, on which our field-theory derivation is based. This is emphasized by the fact that Ref.\ \cite{Bravi_2016} excludes the presence of (non-linear) self-interactions, which in turn do not pose a difficulty to the 2PI approach.}

In comparing our field-theory results for a stochastic process with additive noise and quadratic non-linearity (cf.\ section \ref{subsec:quad_non_lin}) to advanced numerical integrators for SDEs \cite{DifferentialEquations.jl-2017, bezanson2017julia}, we find that only an adaptive integrator having weak convergence \cite{kloeden2013convergence} of order $3.0$ is able to reproduce our results for the second cumulant, albeit at a markedly greater computational expense. Presently, however, there are no algorithms of weak order higher than 3.0 implemented in \cite{DifferentialEquations.jl-2017}, for instance. Thus for non-linear SDEs the direct numerical computation of cumulants of higher order appears to be challenging with current methods. From this perspective, applying the 2PI effective action to non-linear stochastic processes is a first step towards the possible implementation of higher-order effective actions (e.g. four-particle irreducible {\cite{doi:10.1063/1.1704062, vasiliev1998functional, carrington20044pi}}), which would enable the computation of fourth cumulants that should otherwise require a numerical integrator of high weak order or unfeasibly many integration steps.

Leveraging the 2PI effective action as presented in this paper opens up several perspectives for future research. It would be interesting to see whether it is possible to identify models where the 2PI effective action is expected to perform well, while available numerical SDE solvers fail, e.g. SDEs with polynomial non-linearities of higher degree than investigated here. Furthermore, an extension to non-Gaussian initial states is possible in principle {via $n$PI} effective actions (where $n>2$ Legendre transforms are performed with respect to higher moments) \cite{doi:10.1063/1.1704062, vasiliev1998functional, PhysRevD.70.105010}. Evaluating the performance of $n$PI effective actions, if numerically feasible, on stochastic processes with strong non-linearities is another interesting perspective.

\section*{Acknowledgements} The author would like to thank Michael Kajan, Francisco Meirinhos, Peter K. Schuhmacher, and Philipp Knechtges for helpful discussions and comments. Early work on this manuscript was funded in part by the Deutsche Forschungsgemeinschaft (DFG) within the Cooperative Research Center SFB/TR 185 (277625399).

\end{document}